\documentclass[12pt]{article}
\usepackage{amsfonts}
\usepackage{amssymb}
\usepackage{color}
\usepackage{graphicx}

\newcommand{\A}{{\mathfrak A}}

\newcommand{\R}{{\cal R}}
\newcommand{\F}{{\cal F}}
\newcommand{\Sc}{{\cal S}}

\newcommand{\mc}{\mathcal}

\newcommand{\be}{\begin{equation}}
\newcommand{\en}{\end{equation}}
\newcommand{\bea}{\begin{eqnarray}}
\newcommand{\ena}{\end{eqnarray}}
\newcommand{\beano}{\begin{eqnarray*}}
\newcommand{\enano}{\end{eqnarray*}}

\newcommand{\G}{{\cal G}}

\newcommand{\1}{1 \!\! 1}
\newcommand{\ST}{\mc S}

\newcommand{\Hil}{\mc H}

\catcode `\@=11 \@addtoreset{equation}{section}
\def\theequation{\arabic{section}.\arabic{equation}}
\catcode `\@=12
\begin{document}

\begin{center}
{\Large \textbf{A quantum-like view to a generalized two players game}} \vspace{2cm%
}\\[0pt]

{\large F. Bagarello}
\vspace{3mm}\\[0pt]
DEIM, Facolt\`{a} di Ingegneria,\\[0pt]
Universit\`{a} di Palermo, I - 90128 Palermo,  and\\
\, INFN, Sezione di Torino, Italy.\\[0pt]
E-mail: fabio.bagarello@unipa.it\\[0pt]
home page: www.unipa.it/fabio.bagarello \vspace{8mm}\\[0pt]
\end{center}

\vspace*{2cm}

\begin{abstract}
\noindent This paper consider the possibility of using some quantum tools in decision making strategies. In particular, we consider here a dynamical open quantum system helping  two players, $\G_1$ and $\G_2$, to take their decisions in a specific context. We see that, within our approach, the final choices of the players do not depend in general on their initial {\em mental states}, but they are driven essentially  by the environment which interacts with them. The model proposed here also considers interactions of different nature between the two players, and it is simple enough to allow for an analytical solution of the equations of motion.
\end{abstract}

\vfill

\newpage


\section{Introduction}

In recent years the scientific literature has seen a growing interest in the possibility of using quantum ideas and quantum tools in the description of some aspects of several macroscopic systems, systems which, in the common understanding, are usually thought to be {\em purely classical}. This interest has touched very different fields of science, starting with finance, going to ecology, passing through psychology, decision making and so on. The literature is now very rich, and it increases almost every day. We just cite here some recent books, \cite{baa}-\cite{buse}, which cover some of the area mentioned above, but not only.

In this paper we will propose a dynamical approach to a very simple and well known problem in decision making, but in a slightly modified version. Our starting point is what was considered in \cite{khren2}, which is a variation on the theme of the prisoners' dilemma. This is just one of the several contributions existing in the literature related to decision making processes and to brain dynamics, and it is just one of the contributions suggesting the relevance of {\em something quantum} in this kind of problems. For instance, Manousakis in \cite{manou} suggests that the condition describing someone who must still make a choice, could be thought as a superposition of suitable states in a particular Hilbert space, whose coefficients are related to the probabilities of making a particular choice among the various possibilities. He also proposes a time evolution driven by some hyper-simplified hamiltonian. Other {\em effective} hamiltonians are used, in similar contexts, by other authors, \cite{marti,buse2}. Of course these effective  hamiltonians, as such, are usually quite ad hoc and can only be used to describe some particular aspect of the system under analysis.

 Less {\em dynamically oriented} is the paper by Agrawal and Sharda, \cite{agra}, where the authors focus particularly on the probabilistic aspects of the process of decision making. Still other contributions are due, for instance, to Vitiello, Khrennikov et al., \cite{vitiello}, and to Busemeyer et al., \cite{buse3}. In particular, in this last paper the authors confront Markov models of human decision-making with other models somehow connected to quantum mechanics. A common feature of almost all these papers have to do with the probabilistic interpretation of quantum mechanics, where interference effects are quite naturally introduced and described, with respect to what happens using classical ideas, where a similar interpretation is not as natural. Other interesting references are given in \cite{other}.

Let us now go back to the problem we are interested in here. We adopt almost the same notation as in \cite{khren2}. We have two players, $\G_1$ and $\G_2$, and each of them can make two possible choices, "0" and "1". For instance, $0_1$ means that $\G_1$ has chosen "0", while  $1_2$ means that $\G_2$ has chosen "1". More compactly, this choice is indicated as $0_11_2$. In the same way  $1_11_2$ means that both $\G_1$ and $\G_2$ made the same choice, "1". And so on\footnote{Notice that in any choice $X_1Y_2$ the indices 1 and 2 refer to the players, while $X$ and $Y$ refer to the possible choices of the players, 0 or 1.}. Now, let us introduce as in \cite{khren2} four real numbers $a$, $b$, $c$ and $d$ satisfying the inequalities $c>a>d>b$. The payoff of $\G_1$ is $a$ or $c$ if $\G_2$'s choice is $0_2$: $0_10_2$ corresponds to a payoff $a$, while $1_10_2$ corresponds to $c$. On the other hand, if $\G_2$ chooses 1, $1_2$,  $\G_1$'s payoff  is $b$ or $d$: $0_11_2$ corresponds to $b$, while $1_11_2$ corresponds to $d$. The situation is summarized in the following table, \cite{khren2}:\vspace{2mm}

\begin{center}
    \begin{tabular}{ | l | l | l | }
    \hline
    $\G_1\backslash \G_2$ & $0_2$ & $1_2$  \\ \hline
    $0_1$ & $(a\backslash a)$ & $(b\backslash c)$  \\ \hline
    $1_1$ & $(c\backslash b)$ & $(d\backslash d)$  \\ \hline
    \end{tabular}
\end{center}

\vspace{2mm}

This table represents the point of view of both $\G_1$ and $\G_2$. For instance, if the two players choose $0$ ($0_10_2$), they both have a payoff $a$. Analogously, choice $1_11_2$ corresponds to the same payoff $d$ for $\G_1$ and $\G_2$. On the other hand, different choices of the players correspond to different payoffs: the choice $0_11_2$ produces a payoff $b$ for $\G_1$, and $c$ for $\G_2$, while $1_10_2$ produces a payoff $c$ for $\G_1$, and $b$ for $\G_2$. The table shows that if $\G_1$ chooses 1, then he can get the maximum payoff, $c$ (if $\G_2$ chooses 0), or a small one, $d$ (if $\G_2$ chooses 1). Hence this could be the best choice, but it can also have {\em bad} consequences. On the other hand, if $\G_1$ chooses 0, then he can have the minimum payoff, $b$ (if $\G_2$ chooses 1), or a better one, $a$ (if $\G_2$ chooses 0). Hence $\G_1$ should make choice 1, hoping that $\G_2$ makes choice 0, so that he gets the maximum payoff $c$.  Of course, with this choice there exists also the possibility that he gets $d$, which is less than $a$ (corresponding to $0_10_2$). But, for sure, choosing 1, $\G_1$ will not get $b$, which is the lowest possible payoff.  Hence, if what $\G_1$ really hopes is not to get the lowest payoff\footnote{This is what in the literature is called loss-aversion}, he \underline{must}  choose 1. A similar analysis of the table from the point of view of $\G_2$ suggests that, if also $\G_2$ wants to avoid to get the worst payoff, he has to choose 1. Then, if $\G_1$ and $\G_2$ are {\em rational players}, meaning that they are both interested not to get $b$, they should both choose 1 ($1_11_2$). 

In \cite{khren2} this problem has been considered using quantum techniques: a {\em mental state vector} belonging to some suitable Hilbert space is associated to each possible choice of the players, and it is used to describe the situation. In particular, since we have here only four possible choices, the Hilbert space is four dimensional. We will give more details, adapted to our aims, in Section II. The dynamics of the vector is deduced  by a master equation, and the final decision is related to the equilibrium solution of this  equation.

In this paper we consider a similar system from a slightly different point of view, i.e. from the point of view of quantum open systems. In our opinion, this choice is more realistic, since we consider the possibility that the two players interact with {\em the external world}, to make up their mind and to take their decisions. Of course, this is different from the standard version of the two players game, where there is no interaction at all, and this is the reason why we talk of a "similar system". In particular, in our case, the Hilbert space of the model is richer than the one considered in the existing literature. In fact,  most of the papers considering a  quantum approach to decision making deal with finite dimensional Hilbert spaces. This is not the case for us: in our settings, while the players will be {\em attached} to a four-dimensional Hilbert space, the reservoir will not. But, rather than being a problem, in our opinion this makes the structure  more realistic. In fact, the presence of the reservoir mimics well the very many inputs that each player normally takes into account while making  his choice. This is exactly our interpretation of the reservoir: it represents the set of rumors, ideas, suggestions,... coming from the external world and reaching the players. For this reason, the dynamics of the players is provided by an hamiltonian, written following the rules proposed in \cite{bagbook}, which describes not only the two players, but also the reservoir, and, above all, the possible interactions. As we have already said, similar ideas have already been used in the literature on decision making, see \cite{manou,marti,buse2} for instance. However, in these cases, the hamiltonian is often a very simple matrix which, of course, can only be used to describe a particular aspect of the model. On the other hand, our hamiltonian contains rather general information on the system, and the different ingredients of $H$ can be easily identified. Finally, even if, in our knowledge, this is not done in any standard (or quantum) view to the two players game, we will also consider here the possibility of having some interaction between $\G_1$ and $\G_2$, and we will discuss the consequences of this interaction. In particular, we will consider the case in which the two players react {\em in the same way} and the case in which they have {\em opposite reactions}. This will be clarified in the next section. Of course, the presence of this interaction between the players, makes our system even more different from the one considered in \cite{khren2}.

The paper is organized as follows: in the following section we propose the model and we derive its dynamics. Then we consider the cases in which $\G_1$ and $\G_2$ do not interact, and the cases in which they do, and different possibilities are considered. The analysis of the results and our conclusions are discussed in Section III. Finally, to keep the paper self contained, we discuss some important facts in quantum mechanics in the Appendix.

\section{The model and its dynamics}

In this section we will discuss the details of our model, constructing first the vectors of the players, the hamiltonian of the system, and deducing, out of it, the differential equations of motion and their solution, with particular interest to its asymptotic (in time) behavior.

In our game we have two players, $\G_1$ and $\G_2$. Each player could operate two possible choices, 0 and 1. Hence we have four different possibilities, which, following \cite{khren2}, we associate here to four different and mutually orthogonal vectors in a four dimensional Hilbert space $\Hil_\G$. These vectors are $\varphi_{0,0}$, $\varphi_{1,0}$, $\varphi_{0,1}$ and $\varphi_{1,1}$. The first vector, $\varphi_{0,0}$, describes the fact that, at $t=0$, the two players have both chosen 0 ($0_10_2$). Of course, this is not a fixed choice, and can change during the time evolution of the system. Analogously, $\varphi_{0,1}$ describes the fact that, at $t=0$, the first player has chosen 0, while the second has chosen 1 ($0_11_2$). And so on. $\F_\varphi=\{\varphi_{k,l},\,k,l=0,1\}$ is an orthonormal basis for $\Hil_\G$. The general mental state vector of the system $\Sc_\G$ (i.e. of the two players), for $t=0$, is a linear combination
\be
\Psi=\sum_{k,l=0}^1\alpha_{k,l}\varphi_{k,l},
\label{20}\en
where we assume that $\sum_{k,l=0}^1|\alpha_{k,l}|^2=1$ in order to normalize the total probability. Indeed $|\alpha_{0,0}|^2$ is the probability that $\Sc_\G$ is, at $t=0$, in a state $\varphi_{0,0}$, i.e. that both $\G_1$ and $\G_2$ have chosen 0. Notice, incidentally, that $\Psi=\Phi_1\otimes\Phi_2$, where $\Phi_k=x_0^{(k)}\varphi_0^{(k)}+x_1^{(k)}\varphi_1^{(k)}$, $k=1,2$, and where $\alpha_{k,l}$ and $x_j^{(k)}$ are related in an obvious way: $\alpha_{0,0}=x_0^{(1)}x_0^{(2)}$, $\alpha_{1,0}=x_1^{(1)}x_0^{(2)}$, $\alpha_{0,1}=x_0^{(1)}x_1^{(2)}$ and $\alpha_{1,1}=x_1^{(1)}x_1^{(2)}$. We see that the vectors describing $\G_1$ and $\G_2$ are independent, and $\Psi$ is the tensor product of the two.

The first essential difference with respect to what is done in \cite{khren2} is now  the way in which these vectors are constructed: we consider two fermionic operators, see Appendix, i.e. two operators $b_1$ and $b_2$, satisfying the following canonical anti-commutation rules (CAR):
\be
\{b_k,b_l^\dagger\}=\delta_{k,l}\,\1,\qquad \{b_k,b_l\}=0,
\label{21}
\en
where $k,l=0,1$, $\1$ is the identity operator, and $\{x,y\}=xy+yx$. Then we take $\varphi_{0,0}$ as the vacuum of $b_1$ and $b_2$: $b_1\varphi_{0,0}=b_2\varphi_{0,0}=0$, and construct the other vectors out of it:
$$
\varphi_{1,0}=b_1^\dagger\varphi_{0,0}, \quad \varphi_{0,1}=b_2^\dagger\varphi_{0,0}, \quad \varphi_{1,1}=b_1^\dagger\,b_2^\dagger\varphi_{0,0}.
$$
The explicit expressions of these vectors and operators can be found in many textbooks in quantum mechanics, see \cite{rom} for instance: $\varphi_{k,l}=\varphi_k^{(1)}\otimes\varphi_l^{(2)}$, where $\varphi_0=\left(
                                                                          \begin{array}{c}
                                                                            1 \\
                                                                            0 \\
                                                                          \end{array}
                                                                        \right)
$
and
$\varphi_1=\left(
                                                                          \begin{array}{c}
                                                                            0 \\
                                                                            1 \\
                                                                          \end{array}
                                                                        \right)
$. Then,
$$\varphi_{1,0}=\varphi_1^{(1)}\otimes\varphi_0^{(2)}=\left(
                                                                          \begin{array}{c}
                                                                            0 \\
                                                                            1 \\
                                                                          \end{array}
                                                                        \right)\otimes \left(
                                                                          \begin{array}{c}
                                                                            1 \\
                                                                            0 \\
                                                                          \end{array}
                                                                        \right),\qquad
                                                                        \varphi_{1,1}=\varphi_1^{(1)}\otimes\varphi_1^{(2)}=\left(
                                                                          \begin{array}{c}
                                                                            0 \\
                                                                            1 \\
                                                                          \end{array}
                                                                        \right)\otimes \left(
                                                                          \begin{array}{c}
                                                                            0 \\
                                                                            1 \\
                                                                          \end{array}
                                                                        \right),
                                                                        $$
and so on. The matrix form of the operators $b_j$ and $b_j^\dagger$ are also quite simple. For instance,
$$
b_1=\left(
      \begin{array}{cc}
        0 & 1 \\
        0 & 0 \\
      \end{array}
    \right)\otimes \left(
      \begin{array}{cc}
        1 & 0 \\
        0 & 1 \\
      \end{array}
    \right), \qquad b_2=\left(
      \begin{array}{cc}
        1 & 0 \\
        0 & 1 \\
      \end{array}
    \right)\otimes \left(
      \begin{array}{cc}
        0 & 1 \\
        0 & 0 \\
      \end{array}
    \right),
$$
and so on.

Let now $\hat n_j=b_j^\dagger b_j$ be the number operator of the $j$-th player: the CAR above imply that $\hat n_1\varphi_{k,l}=k\varphi_{k,l}$ and $\hat n_2\varphi_{k,l}=l\varphi_{k,l}$, $k,l=0,1$. Then, as already stated, the eigenvalues of these operators correspond to the choice operated by the two players at $t=0$: for instance, $\varphi_{1,0}$ corresponds to the choice $1_10_2$, just because one is the eigenvalue of $\hat n_1$ and zero is the eigenvalue of $\hat n_2$. 

\vspace{2mm}

{\bf Remark:--} One might wonder why, in the description of our model, we use fermionic rather than bosonic operators, as we have done in several other applications in recent years, \cite{bagbook}. This is easily understood, since the eigenvalues of the fermionic number operators $\hat n_j$ are exactly 0 and 1, which are the only possible choices of the players. On the other hand, see \cite{mer}, the eigenvalues of the bosonic number operators  are all the natural numbers (including 0): too many for us!

\vspace{2mm}

Our main effort now consists in {\em giving a dynamics} to  the number operators $\hat n_j$, following the scheme described in \cite{bagbook}. Therefore, what we first need is to introduce a hamiltonian $H$ for the system. Then, we will use this hamiltonian to deduce the dynamics of the number operators as $\hat n_j(t):=e^{iHt}\hat n_j e^{-iHt}$, and finally we will compute the mean values of these operators on some suitable state which is needed to describe, see below, the status of the system at $t=0$. The {\em rules} needed to write down $H$ are described in \cite{bagbook}. The main idea here is that the two players are just part of the {\bf full} system: in order to take their decision, they need to be somehow informed. In fact, it is really the information which creates the final decision. Hence, $\Sc_\G$ must be {\em open}, meaning with this that there must be a reservoir $\R=\R_1\otimes\R_2$, interacting with $\G_1$ and $\G_2$, which is responsible for this sort of information. The reservoir, compared with $\Sc_\G$, is expected to be a very large system since the information is created by several different sources. A possible hamiltonian is therefore the following:
\be
\left\{
\begin{array}{ll}
h=H_{0}+H_{I}, &  \\
H_{0}=\sum_{j=1}^{2}\omega _{j}b_j^\dagger b_j+\sum_{j=1}^{2}\int_{\Bbb R}\Omega_j(k)B_j^\dagger(k)B_j(k)\,dk,   \\
H_{I}=\sum_{j=1}^{2}\lambda_j\int_{\Bbb R}\left(b_j B_j^\dagger(k)+B_j(k)b_j^\dagger\right)\,dk.
\end{array}%
\right.
\label{22}\en
Here $\omega _{j}$ and $\lambda_j$ are real quantities, and $\Omega_j(k)$ are real functions. In analogy with the $b_j$'s, we adopt fermionic operators $B_j(k)$ and $B_j^\dagger(k)$ to describe the reservoir. They depend on $j=1,2$ (two different sub-reservoirs for the two players), and on the real variable\footnote{In principle we should use a discrete variable to label each element of the reservoirs. However, since integrals are quite often easier to be computed than series, as usually done in the literature we consider this label to be real.} $k$, and they satisfy the rules
\be
\{B_i(k),B_l(q)^\dagger\}=\delta_{i,l}\delta(k-q)\,\1,\qquad \{B_i(k),B_j(k)\}=0,
\label{23}
\en
which have to be added to those in (\ref{21}). Moreover each $b_j^\sharp$ anti-commutes with each $B_j^\sharp(k)$: $\{b_j^\sharp, B_l^\sharp(k)\}=0$ for all $j$, $l$ and $k$. Here $X^\sharp$ stands for $X$ or $X^\dagger$.

\subsection{An interlude: why fermionic operators, and why this hamiltonian?}

It may be useful to recall now that, as discussed in the Appendix, the fermionic operators considered above have a very useful characteristic for us: they can be used to construct new self-adjoint operators, the  number operators $\hat n_j=b_j^\dagger b_j$, which are diagonal in the $\varphi_{k,l}$'s, and whose eigenvalues are exactly zero and one. Which, important to stress, are the only two possible choices of our players. As we have already said, this is the core of our choice: we have two main possible choices of the players, and these correspond exactly to two eigenvalues of very simple matrices. Then, as we have already discussed, a rather natural possibility to describe the process of decision making is simply to {\em give a dynamics} to $\hat n_j$. And our claim is that this dynamics is given (in part) by the hamiltonian (\ref{22}). The full hamiltonian is given below, in (\ref{24}).

Let us now concentrate on the meaning of $h$, beginning with the role of the parameters and of the functions. Of course, $\lambda_j$ is an interaction parameter, measuring the strength of the interaction between $\G_j$ and $\R_j$. If, in particular, $\lambda_1=\lambda_2=0$, then $h=H_0$ and, since $[h,\hat n_j]=0$, this would imply that the number operators describing the choices of the two players stay constant in time. In other words, in this case the original choices of $\G_1$ and $\G_2$ are not affected by the time evolution\footnote{Please consider that $h$ is not really the full hamiltonian, see (\ref{24}). In fact, $\hat n_j(t)$  stays really constant in time if $[H,\hat n_j]=0$, which is surely true if $\lambda_1=\lambda_2=0$ and if $\mu_{ex}=\mu_{coop}=0$, see (\ref{24}).}. Both $\omega_j$ and $\Omega_j(k)$ are related to a sort of {\em inertia} of the system, \cite{bagbook}, i.e. to a tendency of a particular part of the system not to change too fast its status. For instance, we will see in Section II.1 that $\Omega_j(k)$ is related to the time needed by $\G_j$ to make his choice. In \cite{bagbook} it is also shown, in several concrete applications, that the values of $\omega_j$ and $\Omega_j(k)$ are related to the magnitude of the oscillations of some relevant functions of the model. For this reason, in analogy with classical mechanics, we adopt the word {\em inertia} in connection with these quantities.

Let us now explain why we have chosen these particular forms of $H_0$, and of $H_I$.

$H_0$ is a sum of diagonal operators, describing the {\em free evolution} of the operators of $\Sc=\Sc_\G\otimes\R$. In fact, for instance, $\sum_{j=1}^{2}\omega _{j}b_j^\dagger b_j=\sum_{j=1}^{2}\omega _{j}\hat n_j$, which is already diagonal in terms of the $\varphi_{k,l}$. Slightly more complicated, but not particularly different, is the part of $H_0$ which refers to the reservoirs: it is also diagonal. Now, suppose that $\G_1$ and $\G_2$ are, at $t=0$ in a definite state, say $\varphi_{0,1}$ (i.e. $0_11_1$), and that the dynamics of the system is {\bf only} given by $H_0$. Then, at $t>0$, the two players will be still described by $\varphi_{0,1}$: no change in their decisions. This is coherent with the fact that, as we have discussed before, $[H_0,\hat n_j]=0$. Stated with different words, we could say that $H_0$ is the simplest quadratic self-adjoint operator in our fermionic operators which commutes with $\hat n_1$ and $\hat n_2$. This ensures that, in absence of interactions, $\G_1$ and $\G_2$ do not change idea.

More interesting is the role of $H_I$. In order to explain its meaning, we have to recall that, see Appendix, $b_j$ and $B_j(k)$ are {\em lowering} operators, while their adjoint $b_j^\dagger$ and $B_j^\dagger(k)$ are {\em raising} operators. For instance, if we consider $b_1\varphi_{1,0}$, we obtain $\varphi_{0,0}$. Then, the action of $b_1$ modify the original choice of the players, $1_10_2$, to the new choice,  $0_10_2$. Similarly, since  $b_1^\dagger\varphi_{0,0}=\varphi_{1,0}$, the action of $b_1^\dagger$ brings $0_10_2$ to $1_10_2$. The operators $B_j(k)$ and $B_j^\dagger(k)$ behave similarly for the reservoir.

Then, it is clear that $H_I$ describes the interaction between the two components of $\R$, $\R_1$ and $\R_2$, with the players: $b_j B_j^\dagger(k)$ describes the fact that, when the amount of information reaching $\G_j$ increases (because of $B_j^\dagger(k)$), $\G_j$ tends to chose 0 (because of $b_j$). On the other hand, $B_j(k)b_j^\dagger$ describes the fact that  $\G_j$  tends to chose 1, when the amount of information reaching him decreases.  Now, recalling that, in our model, what the players really want to avoid is getting the smallest payoff $b$, and recalling that this is achieved by choosing 1, it is natural to interpret the information produced by the reservoir as information of {\em bad quality}: the more it reaches $\G_j$, the more he moves away from his rational choice.

\subsection{Enriching the model}

To make the situation richer and more interesting for us we admit here the possibility that the two players also interact among them and we consider two different possible interactions, by adding a {\em cooperative} and an {\em exchange} effects. The full hamiltonian $H$ is therefore
\be
\left\{
\begin{array}{ll}
H=h+h_{int},   \\
h_{int}=\mu_{ex}\left(b_1^\dagger b_2+b_2^\dagger b_1\right)+\mu_{coop}\left(b_1^\dagger b_2^\dagger+b_2 b_1\right),   \\
\end{array}%
\right.
\label{24}\en
where $\mu_{ex}$ and $\mu_{coop}$ are non negative. In particular, they could be both equal to zero, and in this case $H=h$. In this particular case, $\G_1$ and $\G_2$ do not interact with each other. On the other hand, if $\mu_{ex}\neq0$ and $\mu_{coop}=0$ then  $\G_1$ and $\G_2$ are pushed to make different choices because of the terms $b_1^\dagger b_2$ and $b_2^\dagger b_1$, while they act cooperatively if  $\mu_{ex}=0$ and $\mu_{coop}\neq0$ (because of $b_1^\dagger b_2^\dagger$ and $b_2 b_1$). Finally, we also allow the possibility of having both these contributions, when $\mu_{ex}$ and $\mu_{coop}$ are simultaneously non zero.

Before deducing the time evolution of the relevant observables of the system, it is interesting to discuss the presence, or the absence, of some integrals of motion for the model. In our context, these are (self-adjoint) operators which commute with the hamiltonian. In many concrete situations the existence of these kind of operators gives an hint on how the hamiltonian should look like, \cite{bagbook}, and can be used sometimes to check how realistic our model is. In fcat, this strategy was previously used to fix the form of $H_0$. Let us introduce
\be
N =\sum_{j=1}^2 N_{j}=\sum_{j=1}^2 \left(b_j^\dagger b_j+\int_{\Bbb R}B_j^\dagger(k)B_j(k)\,dk\right),
\label{3add1}\en
with obvious notation.
First of all, it is easy to check that $[N_{j},h]=0$, $j=1,2$, so that $[N,h]=0$. Moreover, even if $\left[N_{j},\mu_{ex}\left(b_1^\dagger b_2+b_2^\dagger b_1\right)\right]\neq0$, we find that $$\left[N,\mu_{ex}\left(b_1^\dagger b_2+b_2^\dagger b_1\right)\right]=0,$$ so that $N$ commutes also with $h+\mu_{ex}\left(b_1^\dagger b_2+b_2^\dagger b_1\right)$. On the other hand, neither $N_{j}$ nor $N$ commute with $\mu_{coop}\left(b_1^\dagger b_2^\dagger+b_2 b_1\right)$ so that, when $\mu_{coop}\neq0$, $N$ ceases to be an integral of motion. This suggests that the cooperation destroys the integral of motion in (\ref{3add1}). This is because the cooperative term in $H$ forces $\G_1$ and $\G_2$ to behave in a similar way forcing, as a consequence, the mean value of $N$ to change with time. On the other hand, if $\mu_{coop}=0$, the creation and the annihilation operators in $H$ always compensate their actions and for this reason $N$ stays constant in time, even if its different contributions in (\ref{3add1}) have a non trivial time evolution.

We can now go back to the analysis of the dynamics of the system. The Heisenberg equations of motion $\dot X(t)=i[H,X(t)]$, see Appendix, can be deduced by using the CAR (\ref{21}) and (\ref{23}) above:
\be
\left\{
\begin{array}{ll}
\dot b_1(t)=-i\omega_1 b_1(t)+i\lambda_1\int_{\Bbb R}B_1(k,t)\,dk-i\mu_{ex}b_2(t)-i\mu_{coop}b_2^\dagger(t),   \\
\dot b_2(t)=-i\omega_2 b_2(t)+i\lambda_2\int_{\Bbb R}B_2(k,t)\,dk-i\mu_{ex}b_1(t)+i\mu_{coop}b_1^\dagger(t),   \\
\dot B_j(k,t)=-i\Omega_j(k) B_j(k,t)+i\lambda_j b_j(t),   \\
\end{array}%
\right.
\label{25}\en
$j=1,2$. The third equation can be rewritten as
$$
B_j(k,t)=B_j(k)e^{-i\Omega_j(k)t}+i\lambda_j\int_0^t b_j(t_1)e^{-i\Omega_j(k)(t-t_1)}\,dt_1
$$
and, taking $\Omega_j(k)=\Omega_j k$, $\Omega_j>0$, standard computations produce
\be
\int_{\Bbb R}B_j(k,t)\,dk=\int_{\Bbb R}B_j(k)e^{-i\Omega_j k t}\,dk+i\pi\frac{\lambda_j}{\Omega_j}\,b_j(t).
\label{26}\en
We refer to \cite{bagbook} for details of this computation and for a discussion on the {\em physical genesis} of this approach.  If we now replace (\ref{26}) in the equations (\ref{25}) for $\dot b_j(t)$, we can write
\be
\dot b(t)=i\,U\,b(t)+i\beta(t),
\label{27}\en
where we have introduced  $\nu_j=i\omega_j+\pi\frac{\lambda_j^2}{\Omega_j}$, $\beta_j(t)=\int_{\Bbb R}B_j(k)e^{-i\Omega_j k t}\,dk$,  $j=1,2$, and
$$
b(t)=\left(
       \begin{array}{c}
         b_1(t) \\
         b_2(t) \\
         b_1^\dagger(t) \\
         b_2^\dagger(t) \\
       \end{array}
     \right), \, \beta(t)=\left(
                               \begin{array}{c}
                                 \lambda_1\beta_1(t) \\
                                 \lambda_2\beta_2(t) \\
                                 -\lambda_1\beta_1^\dagger(t) \\
                                 -\lambda_2\beta_2^\dagger(t) \\
                               \end{array}
                             \right)\, U=\left(
                                              \begin{array}{cccc}
                                                i\nu_1 & -\mu_{ex} & 0 & -\mu_{coop} \\
                                                -\mu_{ex} & i\nu_2 & \mu_{coop} & 0 \\
                                                0 & \mu_{coop} & i\overline{\nu_1} & \mu_{ex} \\
                                                -\mu_{coop} & 0 & \mu_{ex} & i\overline{\nu_2} \\
                                              \end{array}
                                            \right).
$$
The solution of (\ref{27}) is easily found in a matrix form:
\be
b(t)=e^{i\,U\,t}b(0)+i\int_0^t e^{i\,U\,(t-t_1)}\,\beta(t_1)\,dt_1,
\label{28}\en
which is now the starting point for our analysis below.

\subsection{$\G_1$ and $\G_2$ do not interact}

This is {\em almost} the classical two players game, since they do not interact each other, but still both communicate with their environments. As we have discussed before, in this case $\mu_{ex}=\mu_{coop}=0$. Then $U$ is a diagonal matrix, and $e^{i\,U\,t}$ is diagonal as well. Then, from (\ref{28}) we easily deduce that
$$
b_j(t)=e^{-\nu_j t}b_j(0)+i\int_0^t e^{-\nu_j (t-t_1)}\beta_j(t)\,dt_1,
$$
$j=1,2$. From this equation we can obtain $b_j^\dagger(t)$ and, consequently, the number operator $\hat n_j(t)=b_j^\dagger(t)b_j(t)$. However, what is relevant for us is not really $\hat n_j(t)$ itself, but  its mean value on some suitable state on $\Sc$. These states are assumed to be tensor products of vector states for $\Sc_\G$ and states on the reservoir which obey a standard equation, see below. More in details,  for each operator of the form $X_{\Sc}\otimes Y_{\R}$, $X_{\Sc}$ being an operator of $\Sc_\G$ and $Y_{\R}$ an operator of
the reservoir, we consider
$$
\left\langle X_{\Sc}\otimes Y_{\R}\right\rangle :=\left\langle \Psi,X_{\Sc}\Psi\right\rangle \,\omega
_{\R}(Y_{\R}).
$$
Here $\Psi$ is the vector introduced in (\ref{20}), while $\omega _{\R}(.)$ is a state satisfying
the following standard properties, \cite{bagbook}:
\be
\omega _{\R}(1\!\!1_{\R})=1,\quad \omega _{\R}(B_{j}(k))=\omega
_{\R}(B_{j}^{\dagger }(k))=0,\quad \omega _{\R}(B_{j}^{\dagger
}(k)B_{l}(q))=N_{j}\,\delta _{j,l}\delta (k-q),
\label{29}\en
for some constant $N_{j}$.  Also, $\omega
_{\R}(B_{j}(k)B_{l}(q))=0$, for all $j$ and $l$. These formulas for $\omega
_{\R}$ reflect for the reservoir expressions similar to those for $\Sc_\G$. Then
\be
n_j(t)=\left\langle \hat n_j(t)\right\rangle= e^{-2\pi\lambda_j^2/\Omega_j\,t} \|b_j\Psi\|^2+N_j\left(1-e^{-2\pi\lambda_j^2/\Omega_j\,t}\right).
\label{210}\en
What is interesting here is that, if $\lambda_j\neq0$,
\be
n_j(\infty):=\lim_{t\rightarrow\infty}n_j(t)=N_j
\label{211}\en
does not depend on the original {\em state of mind } of the two players, but only on what the reservoir suggests. In fact, independently of the vector $\Psi$ describing probabilistically, at  $t=0$,  the choices of both $\G_1$ and $\G_2$, if $\R_1$ (the part of the reservoir interacting with $\G_1$) has in (\ref{29})  $N_1=0$, then after a sufficiently long time, 0 will be exactly $\G_1$'s choice. On the other hand, if $N_1=1$, then $\G_1$ will eventually choose 1. A similar conclusion can be deduced for $\G_2$. Therefore, when $\G_1$ and $\G_2$ do not interact, their choices are only dictated by their environments. This conclusion looks quite reasonable, in the present context.

\vspace{2mm}

{\bf Remarks:--} (1) Notice that, if $\lambda_j=0$, formula (\ref{210}) reduces to $n_j(t)=\|b_j\Psi\|^2 =n_j(0)$, $\forall\, t$. This is not surprising, since reflects what was already deduced before in absence of interactions of any kind. In this case, in fact, we have seen that the initial state of mind is what really matters for the final decision, since there is no time evolution of the operator $\hat n_j$ at all.

(2) More in general, formula (\ref{210}) suggests the introduction of a sort of {\em characteristic time} for $\G_j$, $\tau_j=\frac{\Omega_j}{2\pi\lambda_j^2}$. The more $t$ approaches $\tau_j$, the bigger the influence of $\R_j$ on $\G_j$ is. In particular, if $\lambda_j\rightarrow 0$, $\tau_j$ diverges. Hence we recover our previous conclusions: $\G_j$ is not influenced at all by $\R_j$, even after a long time. A similar behavior is deduced also when $\Omega_j$ increases: the larger its value, the larger the value of $\tau_j$. In other words, for large $\Omega_j$ the influence of the environment is effective only after a sufficiently long interval. This is not very different from what we have deduced in other systems, \cite{bagbook}, where analogous parameters of the hamiltonian measure the {\em inertia} of that particular part of the system. Of course, $\tau_j$ can be considered as a sort of {\em decision time}.

(3) Since the rational choice of both players is 1, (\ref{211}) shows that rationality really {\em belongs} to $\R_j$, rather than to $\G_j$: in our version of the game, $\G_j$ does not need to be rational, at least if their reservoirs behave rationally!

\subsection{The effect of exchange interaction}

In the following we will fix $\mu_{coop}=0$, allowing $\mu_{ex}$ to be different from zero. In particular, from now on, for concreteness' sake we will work fixing the following values of the other parameters in the hamiltonian: $\omega_1=1$, $\omega_2=2$, $\lambda_1=\lambda_2=0.5$, $\Omega_1=\Omega_2=0.1$. This choice is meant to have almost identical players and  reservoirs. As it is clear, the only difference between $\G_1$ and $\G_2$ is played here by the values of $\omega_1$ and $\omega_2$.

After few computations, calling $V(t)=e^{i\,U\,t}$ and $V_{k,l}(t)$ its $(k,l)$-matrix element,  we deduce that
$$
n_1(t)=|V_{1,1}(t)|^2\|b_1\Psi\|^2+|V_{1,2}(t)|^2\|b_2\Psi\|^2+$$
\be+2\pi\int_0^tdt_1\left[\frac{\lambda_1^2}{\Omega_1}|V_{1,1}(t-t_1)|^2N_1+
\frac{\lambda_2^2}{\Omega_2}|V_{1,2}(t-t_1)|^2N_2\right],
\label{212}\en
and
$$
n_2(t)=|V_{2,1}(t)|^2\|b_1\Psi\|^2+|V_{2,2}(t)|^2\|b_2\Psi\|^2+$$
\be+2\pi\int_0^tdt_1\left[\frac{\lambda_1^2}{\Omega_1}|V_{2,1}(t-t_1)|^2N_1+
\frac{\lambda_2^2}{\Omega_2}|V_{2,2}(t-t_1)|^2N_2\right],
\label{213}\en
To begin with, we consider three different choices for $\mu_{ex}$: (a). $\mu_{ex}=0.01$, (b). $\mu_{ex}=0.05$ and  (c). $\mu_{ex}=0.1$. In all these cases it is possible to check that both $V_{1,1}(t)$ and $V_{1,2}(t)$ converge to zero when $t$ diverges. On the other hand, neither $\int_0^tdt_1|V_{1,1}(t-t_1)|^2$ nor $\int_0^tdt_1|V_{1,2}(t-t_1)|^2$ converge to zero. All these computations can be performed analytically and the explicit result, in case (a), is the following:
\be
\left\{
\begin{array}{ll}
n_1(\infty)\simeq 0.99997 N_1+0.00001 N_2,   \\
n_2(\infty)\simeq 0.00001 N_1+0.99997 N_2.   \\
\end{array}%
\right.
\label{214}\en
How we can see, these are symmetrical, and not very different from the result in (\ref{211}): the two players modify their decision with respect to when $\mu_{ex}=0$, but {\em just a little bit }! This is because $\mu_{ex}$ is too small. In fact, let us consider the case (b) above, $\mu_{ex}=0.05$. In this case, repeating the same computations, we conclude that
\be
\left\{
\begin{array}{ll}
n_1(\infty)\simeq 0.99997 N_1+0.00251 N_2,   \\
n_2(\infty)\simeq 0.00251 N_1+0.99997 N_2,   \\
\end{array}%
\right.
\label{215}\en
which shows that the mixing between $N_1$ and $N_2$ increases a little bit. And, in fact, this mixing increases even more in case (c), when $\mu_{ex}=0.1$: we get
\be
\left\{
\begin{array}{ll}
n_1(\infty)\simeq 0.99039 N_1+0.00958 N_2,   \\
n_2(\infty)\simeq 0.00958 N_1+0.99039 N_2.   \\
\end{array}%
\right.
\label{216}\en

To clarify further the role of the exchange hamiltonian, we now consider much higher values of $\mu_{ex}$, keeping again $\mu_{coop}=0$. Hence we take: (d). $\mu_{ex}=10$ and (e).
$\mu_{ex}=100$. In the first case,  $\mu_{ex}=10$, we find
\be
\left\{
\begin{array}{ll}
n_1(\infty)\simeq 0.59627 N_1+0.40370 N_2,   \\
n_2(\infty)\simeq 0.40370 N_1+0.59627 N_2,   \\
\end{array}%
\right.
\label{217}\en
while in case (e), $\mu_{ex}=100$, we obtain
\be
\left\{
\begin{array}{ll}
n_1(\infty)\simeq 0.50154 N_1+0.49846 N_2,   \\
n_2(\infty)\simeq 0.49846 N_1+0.50154 N_2.   \\
\end{array}%
\right.
\label{218}\en
We believe that, for $\mu_{ex}\gg\mu_{coop}=0$, the two players reach eventually a common choice which should be $n_1(\infty)=n_2(\infty)=\frac{1}{2}(N_1+N_2)$: perfect mixing! Once again, then, the decisions of $\G_1$ and $\G_2$ are driven by the reservoirs but, in this case, the stronger the interaction between $\G_1$ and $\G_2$, the more $\R_1$ and $\R_2$ affect in a symmetric way the two players.

\subsection{The effect of cooperative interaction}

We now consider the case in which only the {\em cooperative} part in the hamiltonian is switched on, $\mu_{coop}\neq0$, while the exchange contribution is turned off, $\mu_{ex}=0$. As before, we will consider, for the same reasons, the following values of the other parameters in the hamiltonian: $\omega_1=1$, $\omega_2=2$, $\lambda_1=\lambda_2=0.5$, $\Omega_1=\Omega_2=0.1$, and then we will put (a). $\mu_{coop}=0.01$, (b). $\mu_{coop}=0.05$ and  (c). $\mu_{coop}=0.1$.

In this case we deduce that
$$
n_1(t)=|V_{1,1}(t)|^2\|b_1\Psi\|^2+|V_{1,4}(t)|^2(1-\|b_2\Psi\|^2)+$$
\be+2\pi\int_0^tdt_1\left[\frac{\lambda_1^2}{\Omega_1}|V_{1,1}(t-t_1)|^2N_1+
\frac{\lambda_2^2}{\Omega_2}|V_{1,2}(t-t_1)|^2(1-N_2)\right],
\label{219}\en
and
$$
n_2(t)=|V_{2,2}(t)|^2\|b_2\Psi\|^2+|V_{2,3}(t)|^2(1-\|b_1\Psi\|^2)+$$
\be+2\pi\int_0^tdt_1\left[\frac{\lambda_1^2}{\Omega_1}|V_{2,3}(t-t_1)|^2(1-N_1)+
\frac{\lambda_2^2}{\Omega_2}|V_{2,2}(t-t_1)|^2N_2\right],
\label{220}\en

Again it is possible to check that all the functions $V_{k,l}(t)$ above converge to zero when $t$ diverges. On the other hand, $\int_0^tdt_1|V_{k,l}(t-t_1)|^2$ admits a non zero limiting value for $t\rightarrow\infty$. The results are the following: in case (a), $\mu_{coop}=0.01$ and $\mu_{ex}=0$, we have
\be
\left\{
\begin{array}{ll}
n_1(\infty)\simeq 0.99997 N_1+0.00001 (1-N_2),   \\
n_2(\infty)\simeq 0.00001 (1-N_1)+0.99997 N_2.
\end{array}%
\right.
\label{221}\en
In case (b), $\mu_{coop}=0.05$ and $\mu_{ex}=0$, we have
\be
\left\{
\begin{array}{ll}
n_1(\infty)\simeq 0.99966 N_1+0.00028 (1-N_2),   \\
n_2(\infty)\simeq 0.00028 (1-N_1)+0.99966 N_2,
\end{array}%
\right.
\label{222}\en
while in case (c), $\mu_{coop}=0.1$ and $\mu_{ex}=0$, we get
\be
\left\{
\begin{array}{ll}
n_1(\infty)\simeq 0.99887 N_1+0.00115 (1-N_2),   \\
n_2(\infty)\simeq 0.00115 (1-N_1)+0.99887 N_2.
\end{array}%
\right.
\label{223}\en
Again we observe that the higher the value of $\mu_{coop}$, the higher the mixing between the effects of the two sub-reservoirs. Hence we are led to formulate a similar conclusion as we did in the previous situation, and we expect that, for $\mu_{coop}\gg\mu_{ex}=0$, the two players arrive to an {\em asymptotic} (in time) choice which is the following: $n_1(\infty)=\frac{1}{2}(N_1+(1-N_2))$, $n_2(\infty)=\frac{1}{2}((1-N_1)+N_2)$.

\subsection{Full hamiltonian}

In this last part, we consider together the effects of the exchange and of the cooperative hamiltonians, still keeping unchanged the values $\omega_1=1$, $\omega_2=2$, $\lambda_1=\lambda_2=0.5$, $\Omega_1=\Omega_2=0.1$. Now both $\mu_{ex}$ and $\mu_{coop}$ will be taken different from zero. In particular, we will consider the situation in which $\mu_{ex}$ and $\mu_{coop}$ are significantly different from each other (which we don't expect is particularly different from what we did before), and the case in which they are similar. More in details, these will be our choices of parameters: Case (a). $\mu_{ex}=0.01$ and $\mu_{coop}=100$; (b). $\mu_{ex}=0.01$ and $\mu_{coop}=1$;  (c). $\mu_{ex}=\mu_{coop}=0.5$; (d). $\mu_{ex}=1$ and  $\mu_{coop}=0.01$ and (e). $\mu_{ex}=100$ and  $\mu_{coop}=0.01$.

In this case we deduce that
$$
n_1(t)=|V_{1,1}(t)|^2\|b_1\Psi\|^2+|V_{1,2}(t)|^2\|b_2\Psi\|^2+|V_{1,3}(t)|^2(1-\|b_1\Psi\|^2)+|V_{1,4}(t)|^2(1-\|b_2\Psi\|^2)+$$
$$+2\pi\int_0^tdt_1\frac{\lambda_1^2}{\Omega_1}\left[|V_{1,1}(t-t_1)|^2N_1+|V_{1,3}(t-t_1)|^2(1-N_1)\right]+$$
\be +2\pi\int_0^tdt_1\frac{\lambda_2^2}{\Omega_2}\left[|V_{1,2}(t-t_1)|^2N_2+|V_{1,4}(t-t_1)|^2(1-N_2)\right]
\label{224}\en
and
$$
n_2(t)=|V_{2,1}(t)|^2\|b_1\Psi\|^2+|V_{2,2}(t)|^2\|b_2\Psi\|^2+|V_{2,3}(t)|^2(1-\|b_1\Psi\|^2)+|V_{2,4}(t)|^2(1-\|b_2\Psi\|^2)+$$
$$+2\pi\int_0^tdt_1\frac{\lambda_1^2}{\Omega_1}\left[|V_{2,1}(t-t_1)|^2N_1+|V_{2,3}(t-t_1)|^2(1-N_1)\right]+$$
\be +2\pi\int_0^tdt_1\frac{\lambda_2^2}{\Omega_2}\left[|V_{2,2}(t-t_1)|^2N_2+|V_{2,4}(t-t_1)|^2(1-N_2)\right].
\label{225}\en
The following are the results we have deduced in the five cases listed above. We have:

Case (a), $\mu_{ex}=0.01$ and $\mu_{coop}=100$:
\be
\left\{
\begin{array}{ll}
n_1(\infty)\simeq 0.50317 N_1+0.49682 (1-N_2),   \\
n_2(\infty)\simeq 0.49682 (1-N_1)+0.50317 N_2.
\end{array}%
\right.
\label{226}\en

Case (b), $\mu_{ex}=0.01$ and $\mu_{coop}=1$:
\be
\left\{
\begin{array}{ll}
n_1(\infty)\simeq 0.91914 N_1+0.08075 (1-N_2),   \\
n_2(\infty)\simeq 0.08075 (1-N_1)+0.91914 N_2.
\end{array}%
\right.
\label{227}\en

Case (c), $\mu_{ex}=0.5$ and $\mu_{coop}=0.5$:
\be
\left\{
\begin{array}{ll}
n_1(\infty)\simeq 0. 85428N_1+0.00626 (1-N_1)+0.11974 N_2+0.01917 (1-N_2)=\\ \qquad\quad=0.84802 N_1+0.10057 N_2+0.02543,   \\
n_2(\infty)\simeq 0.11974 N_1+0.01917 (1-N_1)+0.85428 N_2+0.00626 (1-N_2)=\\ \qquad\quad=0.10057 N_1+0.84802 N_2+0.02543.
\end{array}%
\right.
\label{228}\en
Notice that, in these equations, the first form has been explicitly written simply because, in this way, the different contributions arising from (\ref{224}) and (\ref{225}) can be easily identified.

Case (d), $\mu_{ex}=1$ and $\mu_{coop}=0.01$:
\be
\left\{
\begin{array}{ll}
n_1(\infty)\simeq 0.99210 N_1+0.00795 N_2,   \\
n_2(\infty)\simeq 0.00795 N_1+0.99210 N_2,
\end{array}%
\right.
\label{229}\en
and, finally, Case (e), $\mu_{ex}=100$ and $\mu_{coop}=0.01$:
\be
\left\{
\begin{array}{ll}
n_1(\infty)\simeq 0.50308 N_1+0.49692 N_2,   \\
n_2(\infty)\simeq 0.49692 N_1+0.50308 N_2.
\end{array}%
\right.
\label{230}\en
We postpone our detailed analysis of these and of the previous results to the next section. Here we just want to add that, contrarily to what we have seen in Section II.1, in this more general case we expect that the characteristic time depends also on $\mu_{ex}$ and $\mu_{coop}$, so that these parameters are expected to contribute to the decision time.

\section{Analysis of the results and conclusions}

The first clear output of our analysis suggests that, when $\G_1$ and $\G_2$ do not directly interact, it is really the environment which produces their decisions. Hence the rationality of the players is strongly linked to the nature of the reservoirs: if both reservoirs have $N_j=1$, $j=1,2$, then $n_j(\infty)=1$, and the two players make the most rational choice according to the loss aversion rule. More interesting is the situation when we allow some interaction between $\G_1$ and $\G_2$. In particular our results show that, when at least one between $\mu_{ex}$ or $\mu_{coop}$ is different from zero, and small, the value of $n_j(\infty)$ is essentially decided again by the $j$-th part of the reservoir. However, when the numerical values of one of the two parameters increase, then some {\em mixing} is possible. For instance, we see that when $\mu_{ex}=100$ and $\mu_{coop}=0$, $n_1(\infty)\simeq 0.50154 N_1+0.49846 N_2$ and $n_2(\infty)\simeq 0.49846 N_1+0.50154 N_2$. This means that, even if the two components of the reservoir do not mutually interact, the existence of a direct interaction between $\G_1$ and $\G_2$ {\em mixes the cards}: the final decision of each player is not only related to the value of his own part of reservoir (i.e. to $N_1$ or to $N_2$), but it is a mixture of the two, and, at least for this high value of $\mu_{ex}$, in this mixture $N_1$ and $N_2$ have almost the same weights for  $\G_1$ and $\G_2$. A similar behavior is observed also when $\mu_{ex}=0$ while $\mu_{coop}$ increases: again we have a stronger and stronger mixing of the effects of $\R_1$ and $\R_2$ for $\mu_{coop}$ increasing. However as we see from (\ref{221})-(\ref{223}), $N_1$ mixes with $1-N_2$ (rather than with $N_2$) and $N_2$ with $1-N_1$ (rather than with $N_1$). Hence, this contribution in the hamiltonian, behaves differently from the other one, and this is natural, due to the different kind of the interactions. When we consider both contributions in $h_{int}$, the two effects come together and we see this in formulas (\ref{226})-(\ref{230}). From these formulas we also see that in the extreme situations (when $\mu_{ex}$ is much smaller or much larger than $\mu_{coop}$), not unexpectedly the two final decisions of $\G_1$ and $\G_2$ are similar to the previous cases (i.e. to the cases in which one of the $\mu$'s was zero). On the other hand, when $\mu_{ex}=\mu_{coop}$, the two effects are both clearly visible, see formula (\ref{228}).

In order to compare these results with those in the Introduction, we begin with a very evident fact: the initial state of mind of the players plays absolutely no role in the final decision, except when there is no interaction at all. No matter which was their {\em status} at $t=0$, its effect simply disappears when $t$ increases. This is clearly a measure of the fact that our model is not really the two-player game proposed in \cite{khren2}, as we have already stressed before, but a slightly different version of that.

Let us now consider four different cases, depending on the values of $N_j$ of $\R_j$. Case (I): $N_1=N_2=0$;  Case (II): $N_1=0$ and $N_2=1$;  Case (III): $N_1=1$ and $N_2=0$;  Case (IV): $N_1=N_2=1$. From the formulas of Section II we deduce the following:
\begin{enumerate}

\item the only way in which both $\G_1$ and $\G_2$ choose 1 if when $N_1=N_2=1$, but not for all values of $\mu_{ex}$ and $\mu_{coop}$. For instance, apparently this is not so when $\mu_{coop}\ll\mu_{ex}$. However, when this happens, $n_1(\infty)$ and $n_2(\infty)$ still coincide.

\item on exactly the opposite side, when $N_1=N_2=0$ the values of $n_1(\infty)$ and $n_2(\infty)$ stay always very low, except again when $\mu_{coop}\ll\mu_{ex}$. Even now, when this happens, $n_1(\infty)$ and $n_2(\infty)$ still coincide. The larger $\mu_{coop}$ with respect to $\mu_{ex}$, the bigger the value $n_1(\infty)=n_2(\infty)$ which approaches asymptotically, as our computation suggests, the value $\frac{1}{2}$.

\item when $N_1=0$ and $N_2=1$ in most of the cases considered here $n_1(\infty)$ stays close to 0 while $n_2(\infty)$ is close to 1. However, when $\mu_{ex}\ll\mu_{coop}$, again our numerical results suggest that $n_1(\infty)=n_2(\infty)\simeq\frac{1}{2}$. Specular (and similar) conclusions can be deduced when $N_1=1$ and $N_2=0$.

\item while there is apparently no other way to get $n_1(\infty)=n_2(\infty)=1$ than having $N_1=N_2=1$, there exist several possibilities to have $n_1(\infty)=n_2(\infty)$. Therefore, in a slightly modified version of the game in which we look for equal decisions (not necessarily equal to 1), we have plenty of possibilities in which this happens.

\end{enumerate}

\vspace{2mm}

{\bf Remarks:--}  (1) A different possibility, which may be closer to the usual interpretation of what is {\em quantum} in decision making, is to look at the $n_j(\infty)$ we have deduced before in a probabilistic way. For instance, rather than looking at $n_j(\infty)$ as the {\em real} decision taken by $\G_j$, we could consider it as a sort of probability that $\G_j$ chooses 0 or 1. Then, instead of looking to square modula of the coefficients of the vectors in $\Psi$, we directly look at  $n_j(\infty)$. But this does not fit well with our general interpretation, see \cite{bagbook}, and we will not insist on it here.

(2) It should probably be stressed that the payoffs $a$, $b$, $c$ and $d$ do not enter explicitly in the definition of the hamiltonian, at least in the model considered here. In fact, we are interested here in the possibility that $\G_j$ make the rational choice for a fixed choice of parameters satisfying $c>a>d>b$, whatever this choice is. Changing their values, but maintaining these inequalities, we don't affect the players' behavior, of course. Nevertheless, it could be interesting to look for some different model in which the role of the payoffs is evident in the hamiltonian of the system itself or directly in the state describing the system at $t=0$. The (probably) easiest way to include the payoffs directly in the hamiltonian is to assume, for instance, that $\mu_{ex}$ and $\mu_{coop}$ depend explicitly on $a$, $b$, $c$ and $d$. For instance, if this dependence is such that $\mu_{ex}>\mu_{coop}$, then the effect of the exchange interaction would be stronger than that of the cooperative term in $h_{int}$. More sophisticated dependencies could be considered, like for instance some nonlinear extra term in $H$, depending on the payoffs. But this would make very hard, if not impossible, to get an exact analytical solution, and perturbative expansions should be possibly used.

\vspace{2mm}

This is probably just the beginning of the story: there are still several possible aspects to be considered. First of all, we have considered here just a particular choice of the many parameters of $H$. A natural question is what changes when these parameters, and in particular those which reflect the nature of the players, are fixed in a different way. For instance, in view of the meaning of the $\omega_j$'s we have deduced for other systems, \cite{bagbook}, we could expect a larger {\em inertia} of, say, $\G_1$ with respect to $\G_2$ if $\omega_1\gg\omega_2$: $\G_1$ changes his original idea slowly, when compared to $\G_2$. However, from the point of view of $n_1(\infty)$ and $n_2(\infty)$, we don't expect this will change much our conclusions, but at most some (minor) details, like the decision time. Moreover, the hamiltonian we have considered here is just one among all the possible choices. Indeed, in our opinion, it is a rather natural choice and, when compared with other possibilities, allows a more natural interpretation. Still, one could look for other possibilities, and for instance one could try to add non linearities in the model. However, in this case, numerical techniques should most probably be adopted. Another interesting aspect is the following: is there any other problem in decision making theory in which the method proposed here could be applied? We believe this is very plausible. These are some of the aspects we plan to consider in a close future.

\section*{Acknowledgements}

The author acknowledges partial support from Palermo University and from G.N.F.M. The author also thanks the referees for their suggestions, useful to produce an improved version of the manuscript.

\renewcommand{\theequation}{A.\arabic{equation}}

\section*{Appendix:  Few results on the number representation}
To keep the paper self-contained, we discuss here few important facts in quantum mechanics and in the so--called number representation. More details can be found, for instance, in \cite{mer,rom}, as well as in\cite{bagbook}.

Let $\Hil$ be an Hilbert space, and $B(\Hil)$ the set of all the (bounded) operators on $\Hil$.    Let $\ST$ be our physical system, and $\A$ the
set of all the operators useful for a complete description of $\ST$, which includes the \emph{observables} of $\ST$. For simplicity, it is
convenient (but not really necessary) to assume that  $\A$ coincides with $B(\Hil)$ itself. The description of the time evolution of $\ST$ is related to a self--adjoint
operator $H=H^\dagger$ which is called the \emph{Hamiltonian} of $\ST$, and which in standard quantum mechanics represents  the energy of
$\ST$. In this paper we have adopted the so--called \emph{Heisenberg} representation, in which the time evolution of an observable $X\in\A$ is given by
\be X(t)=\exp(iHt)X\exp(-iHt), \label{a1} \en or, equivalently, by the solution of the differential equation \be
\frac{dX(t)}{dt}=i\exp(iHt)[H,X]\exp(-iHt)=i[H,X(t)],\label{a2} \en where $[A,B]:=AB-BA$ is the \emph{commutator} between $A$ and $B$. The time
evolution defined in this way is a one--parameter group of automorphisms of $\A$.

An operator $Z\in\A$ is a \emph{constant of motion} if it commutes with $H$. Indeed, in this case, equation (\ref{a2}) implies that $\dot
Z(t)=0$, so that $Z(t)=Z$ for all $t$.

In some previous applications, \cite{bagbook}, a special role was played by the so--called \emph{canonical commutation
relations}. Here, these are replaced by the so--called \emph{canonical anti--commutation relations} (CAR): we say that a set of operators
$\{a_\ell,\,a_\ell^\dagger, \ell=1,2,\ldots,L\}$ satisfy the CAR if the conditions \be \{a_\ell,a_n^\dagger\}=\delta_{\ell n}\1,\hspace{8mm}
\{a_\ell,a_n\}=\{a_\ell^\dagger,a_n^\dagger\}=0 \label{a3} \en hold true for all $\ell,n=1,2,\ldots,L$. Here, $\1$ is the identity operator
and $\{x,y\}:=xy+yx$ is the {\em anticommutator} of $x$ and $y$. These operators, which are widely analyzed in any textbook about quantum
mechanics (see,  for instance, \cite{mer,rom}) are those which are used to describe $L$ different \emph{modes} of fermions. From these
operators we can construct $\hat n_\ell=a_\ell^\dagger a_\ell$ and $\hat N=\sum_{\ell=1}^L \hat n_\ell$, which are both self--adjoint. In
particular, $\hat n_\ell$ is the \emph{number operator} for the $\ell$--th mode, while $\hat N$ is the \emph{number operator of $\ST$}.
Compared with bosonic operators, the operators introduced here satisfy a very important feature: if we try to square them (or to rise to higher
powers), we simply get zero: for instance, from (\ref{a3}), we have $a_{\ell}^2=0$. This is related to the fact that fermions satisfy the Fermi
exclusion principle \cite{rom}.

The Hilbert space of our system is constructed as follows: we introduce the \emph{vacuum} of the theory, that is a vector $\varphi_{\bf 0}$
which is annihilated by all the operators $a_\ell$: $a_\ell\varphi_{\bf 0}=0$ for all $\ell=1,2,\ldots,L$. Such a non zero vector surely exists. Then we act on $\varphi_{\bf 0}$
with the  operators $a_\ell^\dagger$ (but not with higher powers, since these powers are simply zero!): \be
\varphi_{n_1,n_2,\ldots,n_L}:=(a_1^\dagger)^{n_1}(a_2^\dagger)^{n_2}\cdots (a_L^\dagger)^{n_L}\varphi_{\bf 0}, \label{a4} \en $n_\ell=0,1$ for
all $\ell$. These vectors form an orthonormal set and are eigenstates of both $\hat n_\ell$ and $\hat N$: $\hat
n_\ell\varphi_{n_1,n_2,\ldots,n_L}=n_\ell\varphi_{n_1,n_2,\ldots,n_L}$ and $\hat N\varphi_{n_1,n_2,\ldots,n_L}=N\varphi_{n_1,n_2,\ldots,n_L},$
where $N=\sum_{\ell=1}^Ln_\ell$. Moreover, using the  CAR, we deduce that $$\hat
n_\ell\left(a_\ell\varphi_{n_1,n_2,\ldots,n_L}\right)=(n_\ell-1)(a_\ell\varphi_{n_1,n_2,\ldots,n_L})$$ and $$\hat
n_\ell\left(a_\ell^\dagger\varphi_{n_1,n_2,\ldots,n_L}\right)=(n_\ell+1)(a_l^\dagger\varphi_{n_1,n_2,\ldots,n_L}),$$ for all $\ell$. Then
 $a_\ell$ and $a_\ell^\dagger$ are  called the
\emph{annihilation} and the \emph{creation} operators. Notice that, in some sense, $a_\ell^\dagger$ is {\bf also} an annihilation operator since,
acting on a state with $n_\ell=1$, we destroy that state.

The Hilbert space $\Hil$ is obtained by taking  the linear span of all these vectors. Of course, $\Hil$ has a finite dimension. In particular,
for just one mode of fermions, $dim(\Hil)=2$. This also implies that, contrarily to what happens for bosons, all the fermionic operators are
bounded.

The vector $\varphi_{n_1,n_2,\ldots,n_L}$ in (\ref{a4}) defines a \emph{vector (or number) state } over the algebra $\A$  as \be
\omega_{n_1,n_2,\ldots,n_L}(X)= \langle\varphi_{n_1,n_2,\ldots,n_L},X\varphi_{n_1,n_2,\ldots,n_L}\rangle, \label{a5} \en where
$\langle\,,\,\rangle$ is the scalar product in  $\Hil$. As we have discussed in \cite{bagbook}, these states are useful
to \emph{project} from quantum to classical dynamics and to fix the initial conditions of the considered system.

\end{document}